\documentstyle[aps,multicol,epsf]{revtex}
\begin{document}
\preprint{}
\draft
\title{Localization of Two Interacting Particles in One-Dimensional Random
Potential}
\author{P. H. Song and Doochul Kim}
\address{Department of Physics and Center for Theoretical Physics, Seoul 
National University, Seoul 151--742, Korea}
\date{\today}
%\ \hspace{5in}{\normalsize SNUTP }\\
\maketitle
\begin{abstract}
We investigate the localization of two interacting particles in 
one-dimensional random potential.  Our definition of the two-particle 
localization length, $\xi$, is the same as that of v. Oppen {\it et al.}
[Phys. Rev. Lett. {\bf 76}, 491 (1996)] and $\xi$'s for chains of finite
lengths are calculated numerically using the recursive Green's function 
method for several values of the strength of the disorder, $W$, and 
the strength of interaction, $U$.  When $U=0$, $\xi$ approaches a value 
larger than half the single-particle localization length as the system 
size tends to infinity and behaves as $\xi \sim W^{-\nu_0}$ for small $W$ 
with $\nu_0 = 2.1 \pm 0.1$.  When $U\neq 0$, we use the finite size 
scaling ansatz and find the relation $\xi \sim W^{-\nu}$ with $\nu = 
2.9 \pm 0.2$.  Moreover, data show the scaling behavior $\xi \sim 
W^{-\nu_0} g(b |U|/W^\Delta)$ with $\Delta = 4.0 \pm 0.5$.  

\end{abstract}
$$\\$$


\begin{multicols}{2}
\narrowtext

Recently, there has been intensive attention
\cite{she,imr,fra1,wei1,opp,wei2,rom,fra2,fra3,voj} focused on the problem 
of the 
localization of two interacting particles in one-dimensional (1D) random 
potential.  With a few assumptions on the statistical nature of
single-particle localized states, Shepelyansky\cite{she} has mapped the problem
approximately to a random band matrix model and obtained an expression
for the two-particle localization length, $\xi$, as
\begin{equation}
\xi \simeq U^2 \frac{\xi_1^2}{32},
\end{equation}
where $U$ is the on-site interaction in unit of the hopping
energy between nearest neighbor pair sites, and $\xi_1$ the single-particle
localization length.  This expression is surprising because it implies
that $\xi$ can exceed $\xi_1$ at
sufficiently small disorder, i.e. sufficiently large $\xi_1$.  Later
Imry\cite{imr} has provided a support for Eq. (1) by invoking the Thouless 
scaling
argument.  However, the methods employed in Refs.~[1] and [2] are partly
approximate and the strict validity of the expression of Eq.~(1) is
questionable as discussed in, e.g. Refs.~[3-8] and [10].

Many authors\cite{fra1,wei1,opp,wei2} have tried to find more 
refined expressions than Eq.~(1)
by improving the assumptions of Shepelyansky.  However, at this stage, 
there exist controversies yet as to the quantitative expression for 
$\xi$ like Eq.~(1).  Frahm {\it et al.}\cite{fra1} obtained the relation 
$\xi \sim \xi_1^{1.65}$ by the transfer matrix method while an approximate
calculation of Green function by v. Oppen {\it et al.}\cite{opp} leads to the
hypothesis $\xi = \xi_1/2 + c |U| \xi_1^2$, where $c$ is a constant
depending on the statistics of the particles.  With the assumption
that the level statistics of two interacting particles is described by
a Gaussian matrix ensemble, Weinmann and Pichard\cite{wei1} argued that 
$\xi$ increases initially as $|U|$ before eventually behaving as $U^2$.  
Moreover, very recently, R\"{o}mer and Schreiber have claimed the 
disappearance of the enhancement as the system size grows (see Refs.~[7] 
and [8]).

Some of these discrepancies, especially between numerical studies, 
are due to different definitions for two-particle localization length 
between authors and also to lack of careful analysis of the finite size
effect of the system size.  The system under study is a ``quantum
mechanical two-body problem" in a sense.  Motion of the two particles
can be decomposed into the motion of the center of mass (CM) and that
of the relative coordinate.  We are interested in the CM motion since
the wavefunction describing the relative motion would not be
different from that arising from the single-particle localization
problem in the thermodynamic limit if the interaction is short-ranged.
Therefore, in this paper, we use the same definition for $\xi$ as
introduced by v. Oppen~{\it et al.}\cite{opp} for the measure for 
localization length of the CM:
\begin{equation}
\frac{1}{\xi} = -\lim_{|n-m| \rightarrow \infty} \frac{1}{|n-m|} \ln
|\langle n,n|G|m,m\rangle |.
\end{equation}
Here, $G$ is the Green function and $|i,j\rangle$ is a two-particle
state in which the particle 1 (2) is localized at a site $i$ $(j)$.
The above definition is reasonable for a description of the CM motion
as long as $U$ is smaller than or of the order of the hopping energy
between sites\cite{opp}.  In practice, we calculate $\xi_N$ defined 
below in Eq.~(4) for chains of finite lengths {\it without any approximation} 
for
several values of $W$ and $U$.  We then estimate $\xi$ by extrapolating
$\xi_N$ using the finite size scaling ansatz.  When $U=0$, we find $\xi \sim
W^{-\nu_0}$ with $\nu_0 = 2.1 \pm 0.1$.  Data for $U \neq 0$ lead to the
relation $\xi \sim W^{-\nu}$ with $\nu = 2.9 \pm 0.2$.  Also the data
lead us to propose a scaling form $\xi \sim W^{-\nu_0} g(b|U|/W^\Delta)$,
where $g(y)$ is a scaling function with the property $g(y \rightarrow 0)
=$ constant and $g(y \rightarrow \infty) \sim y^{(\nu-\nu_0)/\Delta}$.
$\Delta$ is given as $4.0 \pm 0.5$.

We work within the tight-binding equation given by 
\begin{eqnarray}
\psi_{m+1,n} + \psi_{m-1,n} + \psi_{m,n+1} + \psi_{m,n-1} \nonumber \\
= (E-\epsilon_m-\epsilon_n-U \delta_{m,n})\psi_{m,n},
\end{eqnarray}
where $\psi_{i,j} = \langle i,j|\psi \rangle$, $E$ is the energy of the
two particles and $\delta_{m,n}$ the Kronecker delta.  $m$ and $n$ are
the site indices of a chain of length $N$ and range from 1 to $N$,
$\epsilon_m$ is the random site energy chosen from a box
distribution with interval $[-W/2,W/2]$\cite{com1} and the hard wall boundary
condition, i.e. $\psi_{0,n} = 0$ and so on, is used.  As was previously
noted\cite{fra1,rom}, if one interprets $(m,n)$ as Cartesian coordinates 
of a square
lattice of size $N \times N$, the Hamiltonian describes a single particle
in a two-dimensional random potential.  In Eq.~(2), the thermodynamic
limit is first taken and then the limit $|n-m| \rightarrow \infty$.  To
estimate this quantity, we define a sequence $\xi_N$ as
\begin{equation}
\frac{1}{\xi_N} = - \ll\frac{1}{N-1} \ln |\langle 1,1|G_N|N,N\rangle |\gg,
\end{equation}
where $G_N$ represents the Green function for a chain of length $N$ and
the double brackets represent the configurational average.  To be
specific, calculation of $G_N$ amounts to evaluation of the inverse of
the matrix, $(E-\cal{H})$, the size of which is $N^2 \times N^2$.  One
can calculate several elements of $G_N$, i.e. the elements involving the
sites of two opposite edges of the square lattice, very efficiently using
the recursive algorithm of MacKinnon and Kramer\cite{mac}.  We assume 
that $\xi_N$ approaches $\xi$ as $N \rightarrow \infty$.

The on-site interaction of the Hamiltonian given by Eq.~(3) is
relevant only to the spatially symmetric states, which would be
realized, say, for a pair of electrons with total spin zero.  One can 
easily see that the contributions to Eq.~(2) are only from the
spatially symmetric states from following consideration.  The Green's
function represents the transition amplitude from an initial state to
a final state and since the Hamiltonian, Eq.~(3), is invariant
under the exchange operation of two particles, the parity of the
wavefunction is conserved during the time evolution.  Since
the initial state of Eq.~(2) is a doubly occupied state, i.e. a spatially
symmetric state, we are treating only the contributions from symmetric states.

Numerical calculations of $\xi_N$ for various values of $W, U$ and $N$ are
performed for $E=0$ without approximation.  $N$ is varied within the
range $10 \leq N \leq 200$ and for a given parameter set,
configurational average is performed over sufficiently many different
realizations to control the uncertainties of $\xi_N$ within 1\%.

We first examine the case of $U=0$, i.e. the noninteracting two
particles.  In this case, when the total energy of the system is fixed to
$E$, the two-particle wavefunction is a superposition of the products of
two single-particle states of energy $E'$ and $E-E'$, and the Green
function is given by the convolution of two single-particle Green
functions as 
\begin{equation}
\langle i,i|G(E)|j,j\rangle \sim \int dE' \langle
i|G_0(E')|j\rangle\langle i|G_0(E-E')|j\rangle.
\end{equation}
It is a nontrivial problem to calculate $\xi(U=0)$ since there exist
contributions from various energies.  Some authors\cite{wei1,opp} have 
assumed the
relation $\xi(U=0) = \xi_1/2$, i.e. half the single-particle localization
length, which should be, however, seriously examined.  Our numerical data
presented in Fig.~1 show that the assumption is not strictly valid.  The
filled symbols on the $N=\infty$ axis represent $\xi_1/2$ calculated from
the expression $\xi_1 \simeq 105/W^2$\cite{pic} while the empty symbols are 
our
numerical results for $\xi_N$.  Taking into account the fact that the
uncertainty of each data point is smaller than the symbol size, $\xi_N$ does
not seem to extrapolate to $\xi_1/2$ as $N$ tends to infinity.  Moreover,
the discrepancy between the two quantities becomes larger as $W$ gets
smaller.  Therefore, we conclude that within the definition of Eq.~(2),
the single-particle localization length is not an adequate parameter, if
it is qualitatively, for a quantitative description of two-particle
localization problem.  From the data of $N=200$, we get $\xi(U=0) \simeq
70/W^{\nu_0}$ with $\nu_0 = 2.1 \pm 0.1$.

Next, we discuss the case of $U \neq 0$.  Figure 2(a) shows the results 
for $U=1.0$ and $W$ ranging from 0.5 to 10.0.
The $y$ axis label represents the renormalized localization length, i.e.
$\xi_N$ divided by the system size.  For larger values of $W$ and $N$, 
$\xi_N/N$ behaves as $\sim 1/N$, which implies the convergence of
$\xi_N$'s to their constant limiting values.  This means that the
condition $N \gg \xi$ is well satisfied for these data.  However, for
smaller values of $W$, i.e. for $W$ ranging from 0.5 to 1.5, it is not easy to
deduce the value of $\xi$ since $\xi_N$'s increase steadily within the
range of $N$ presented.  Therefore we rely on the scaling idea\cite{mac}, 
which states that $\xi_N/N$ is given by a function of a single parameter, 
i.e. $N/\xi$:
\begin{equation}
\xi_N/N = f(N/\xi).
\end{equation}
The implication of Eq.~(6) is that on a log-log plot all data points of
Fig.~2(a) fall on a single curve when translated by $\ln \xi(W)$ along
the $x$ axis.  As a result, $\xi(W)$'s can be obtained as fitting
parameters.  The result of data collapsing is shown in Fig.~2(b) for the
data set $N \geq 50$.  $\xi(W=5.0)$ has been obtained to be $2.87 \pm 0.01$
by fitting the data set for $W=5.0$ and $N \geq 50$ to the formula $\xi_N
= \xi -A/N$\cite{com2}, where $A$ is a constant.  Other remaining values of
$\xi(W)$'s are obtained by examining the amount of relative translations with
respect to the data set of $W=5.0$.  The scaling plot is quite good and
one can see that the scaling function $f(x)$ behaves as
\begin{equation}
f(x) \sim \left\{ \begin{array}{ll}
1/\sqrt{x} & \mbox{\ \ \ \ if $x \ll 1$}, \\
1/x & \mbox{\ \ \ \ if $x \gg 1$}.
\end{array}
\right.
\end{equation}     
As was previously mentioned, the asymptotic behavior for $x \gg 1$
represents the convergence of $\xi_N$'s to $\xi$.  On the other hand, the
behavior for $x \ll 1$ is very interesting since the same asymptotic
behavior has been found for noninteracting disordered 1D systems\cite{pic}.
For the noninteracting case, the resistance $\rho^0_N$ of a chain of
length $N$ is related to the single-particle localization length
$\xi^0_N$ as\cite{pic} 
\begin{equation}
\rho^0_N = [\cosh(2N/\xi^0_N)-1]/2.
\end{equation}
For $N/\xi^0_N \ll 1$, the right hand side of Eq.~(8) reduces to $\sim
(N/\xi^0_N)^2 \sim N/\xi_1$.  Therefore for the noninteracting case, the
asymptotic behavior for $N \ll \xi_1$ represents the metallic behavior
of the resistance, i.e. the linear increase of the resistance as the chain
length in the metallic regime.  Though no explicit expression like Eq.~(8) 
is not present for the system under study in this paper, we believe
that the same asymptotic behavior for the two cases found here is a
strong indication that the definition in Eq.~(2) is a physically
reasonable one.

The $\xi$'s thus obtained as a function of $W$ are plotted in 
Fig.~2(c)\cite{com3}.
For $0.5 \leq W \leq 5.0$ they are reasonably well fitted to the form of
$\sim W^{-\nu}$, where $\nu$ is given by $2.9 \pm 0.1$.  Within the
error, this value for $\nu$ is different from $\nu_0 = 2.1 \pm 0.1$, i.e.
the critical exponent for $U=0$, and from 4.0, which is the value
expected by Eq. (1).

Further calculations and similar scaling analyses have been performed for
other values of $U$, i.e. 0.2, 0.5, 0.7 and 1.5 upto system size
$N=200$.  It is difficult to determine $\xi$ for $W < 1.5$ and $U <
1.0$ since the corresponding data of $\xi_N$'s do not show scaling
behaviors due to the fact that sufficiently large system sizes have not
been reached for these parameters.  The resulting $\xi$'s (for 1.5 $\leq
W \leq 5.0$ if $U < 1.0$ and for 0.7 $\leq W \leq 5.0$ if $U>1.0$) give
$\nu = 2.7, 3.0, 2.9$ and 3.1 for $U = 0.2, 0.5, 0.7$ and 1.5,
respectively.  Since we do not expect that $\nu$ depends on $U$, we
interpret the variations of the values for $\nu$ as resulting from
numerical uncertainties.  Therefore our final result for the critical
exponent of $\xi$ is $2.9 \pm 0.2$.  

Our result for $\nu$ implies that introduction of nonzero $U$
changes the critical behavior of $\xi$ and, in analogy with thermal 
critical phenomena, the point $W=U=0$ may be regarded as a multicritical 
point and the line $W=0$ as a critical line in the $W-U$ plane.
Then, one may assume a scaling form for $\xi$ as follows;
\begin{equation}
\xi = W^{-\nu_0} g(b|U|/W^\Delta),
\end{equation}
where $g(y)$ is a scaling function, $\Delta$ a crossover exponent and
$b$ a constant.  Here, we used the fact that Eq.~(3) is symmetric 
for $E=0$ so that
$\xi$ depends only on the absolute value of $U$.  The scaling function
should satisfy $g(y \rightarrow 0) =$ constant and $g(y \rightarrow
\infty) \sim y^{(\nu-\nu_0)/\Delta}$ for consistency.  We obtain reasonably 
good scaling plots within the range $\Delta = 4.0 \pm 0.5$.  The
scaling plot for $\Delta = 4.0$ is shown in Fig.~3, where $\xi
W^{2.1}$ is plotted against $U/W^4$ for various values of $W$ and $U$.
Although the data for $W < 1.5$ may appear to deviate from the scaling
curve, taking into account rather large numerical uncertainties
of these data, one can expect that they are consistent with the
scaling behavior of other data points.  We expect
that the crossover between the two asymptotic behaviors occurs at $y
\sim 1$ so that the constant $b$ is estimated to be of the order of
100 from Fig.~3.  Data within the range $0.01 <
U/W^\Delta < 1.0$ ($1 < y < 100$) approximately obey the 
form $\simeq y^{0.23}$, which is shown as a straight line.  Since we
expect the asymptotic behavior $\sim y^{(\nu-\nu_0)/\Delta}$ for this
regime, we obtain $(\nu-\nu_0)/\Delta \simeq 0.23$, i.e. $\nu \simeq 3.0$,
which is in good agreement with our previous estimate, i.e. $\nu = 2.9
\pm 0.2$.  At this stage, we have not found a physical mechanism
regarding the scaling parameter, $U/W^\Delta$ with $\Delta = 4.0
\pm 0.5$.  The quantity $\xi(U)-\xi(U=0)$ might be also of interest.
Our data show that this is consistent with a form $\xi(U)-\xi(U=0)
\sim W^{-2.1} (U/W^4)^{1/2}$ in the region $bU/W^\Delta < 1.0$. 
This behavior shows that the first correction to $g(y)$ 
for $y \ll 1$ is given as $\sim y^{1/2}$.   
On the other hand, our result confirms the enhancement of two
localization length due to the interaction.  The arrow of Fig.~3
represents the value of $\xi W^{\nu_0}$ for $U = 0$ so that it is clearly
seen that $\xi$ monotonically increases with respect to $U$.

Finally, we point out differences between our work and some of those
previously reported.  References [3] and [7] deal with exactly the
same system as ours but use a different definition for the two-particle 
localization length.  As mentioned before, the problem can be considered 
as that of
a noninteracting single particle in a two-dimensional potential.  These
authors study the evolution of a state along one of the edges of
the square lattice.  However, in this paper, we are concerned with the
``pair propagation", more generally, the propagation of the CM of the
two particles.  The definitions for $\xi$ given by Eqs.~(2) and (4) 
describes the propagation along one of the diagonals of the square 
lattice, instead of that along the edge.  Our definition of $\xi$ 
is exactly the same as that of
v. Oppen {\it et al.}\cite{opp}.  However it should be noted that in 
their work, calculation of $\xi$ involves an approximation; the
approximation scheme used in Ref.~[5] fails for small values of $U$
while our results are valid for all values of $U$.

In summary, we have investigated numerically the localization of two
interacting particles in 1D random potential using the definition
introduced previously for the two-particle localization length.  While
we find the enhancement of $\xi$ by the interaction, critical properties
of $\xi$ are different from those reported in previous studies.  We
ascribe the differences to the approximation used in one case and, in 
the other cases, to different definition of $\xi$.
Further works are needed to connect the resistance and the
two-particle localization length and to elucidate the relation between
$\xi$ and $\xi_1$.

This work has been supported by the KOSEF through the CTP and by the 
Ministry of Education through BSRI both at Seoul National University. 
We also thank SNU Computer Center for the computing times on SP2.

\pagebreak

\begin{figure}
\centerline{\epsfxsize=8cm \epsfbox{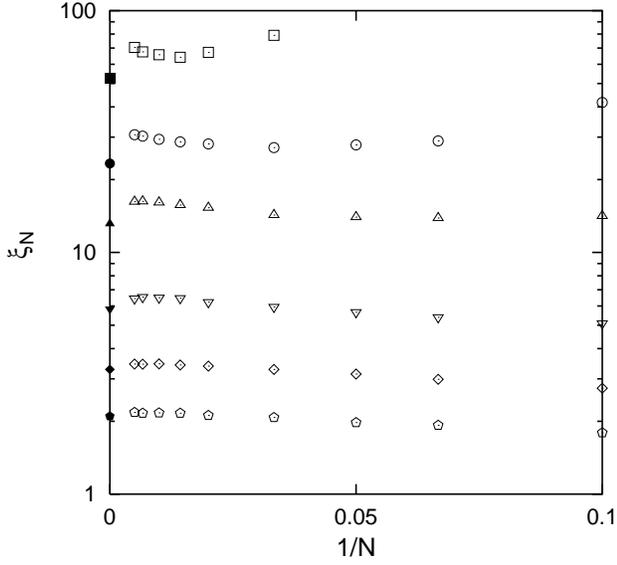}}
\vspace{4mm}
\caption{
 $\xi_N$ (open symbols) as a function of $1/N$ for $U=0$
and $\xi_1/2$ (filled symbols) calculated from the expression $\xi_1
\simeq 105/W^2$: $W = 1.0$ (box), 1.5 (circle), 2.0 (uptriangle), 3.0
(downtriangle), 4.0 (diamond) and 5.0 (pentagon), from top to bottom.
The uncertainty of each data point is less than the symbol size.
}
\end{figure}
\noindent
\setcounter{figure}{2}
\vspace{-0.7cm}
\begin{figure}
\centerline{\epsfxsize=8cm \epsfbox{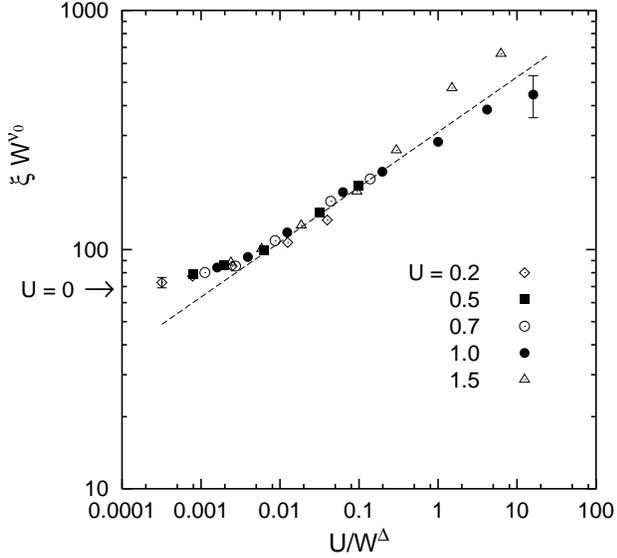}}
\vspace{4mm}
\caption{
The scaling plot of $\xi W^{\nu_0}$ versus
$U/W^\Delta$ with $\nu_0 = 2.1$ and $\Delta = 4.0$.  The typical
uncertainty of data for $U/W^\Delta \geq 1.0$ ($W < 1.5$) is shown
at the rightmost data point, while that for $U/W^\Delta < 1.0$
($W \geq 1.5$) is shown at the leftmost data point.  The straight
line is $\sim x^{0.23}$ and the arrow represents $\xi W^{\nu_0} \simeq
70$ for $U=0$.
}
\end{figure}
\noindent
\setcounter{figure}{1}
\begin{figure}
\centerline{\epsfxsize=8cm \epsfbox{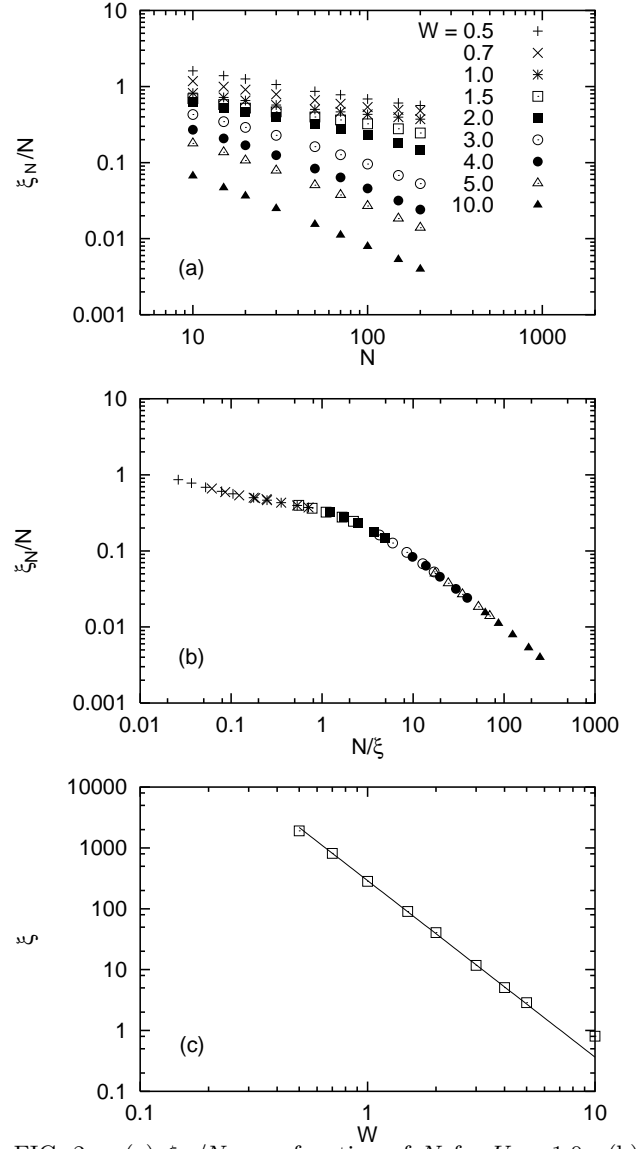}}
\caption{
 (a) $\xi_N/N$ as a function of $N$ for $U = 1.0$.  (b)
Scaling plot constructed from the data of (a) for $N \geq 50$.
$\xi(W)$'s are obtained as fitting parameters by this procedure.  (c)
$\xi$ as a function of $W$.  The data for $W\leq 5.0$ fit well to a
straight line of $\sim W^{-2.9}$ as shown.
}
\end{figure}
\end{multicols}

\begin{references}
\bibitem{com1}
Note that the definition of $W$ in some previous studies, i.e. 
Refs.~[1], [3], [4] and [6], is different from ours by a factor of 2.
\bibitem{com2}
Although this functional form is not theoretically based, the data fit
quite well to this form.  Therefore no significant error in our
results is expected by using this form.
\bibitem{com3}
The numerical uncertainty in $\xi$ is somewhat larger for $W < 1.5$
($\delta(\ln\xi) \sim \delta\xi/\xi \sim 0.2$) than that for $W \geq 1.5$
$(\delta(\ln\xi) \sim 0.05)$.  This is because the data for $W < 1.5$ lie
in a rather flat range, as shown in Fig.~2(b), so that the scaling
plot obtained by translations remains reasonably good within a
comparatively broad range of $\xi$.  Nevertheless, for Fig.~2(c),
uncertainty of each data point is less than the symbol size.
\end{references}
\end{document}